\documentclass[twocolumn,trackchanges]{aastex62}
\usepackage{amsmath,amssymb}
\usepackage{lineno}
\submitjournal{ApJL}

\shorttitle{A consistency test at the epoch of recombination using DESI and Planck measurements}
\shortauthors{Pogosian, Zhao \& Jedamzik}

\hypersetup{
    colorlinks=true,
    linkcolor=blue,
    citecolor=blue,
}

\newcommand{\remove}[1]{}

\def\be{\begin{equation}}
\def\ee{\end{equation}}
\def\ba{\begin{eqnarray}}
\def\ea{\end{eqnarray}}

\def\rd{r_{\rm d}}
\def\ts{\theta_\star}
\def\rstar{r_\star}
\def\rdh{r_{\rm d}h}
\def\om{\Omega_{\rm m}h^2}
\def\omh3{\Omega_{\rm m}h^3}
\def\Om{\Omega_{\rm m}}
\def\sdss{SDSS$^+$}
\def\sn{P$^+$SN}
\def\lcdm{$\Lambda$CDM}
\def\bs{\beta_\star}

\begin{document}

\title{A consistency test of the cosmological model at the epoch of recombination using DESI BAO and Planck measurements}

\author{Levon Pogosian}\thanks{E-mail: \url{levon@sfu.ca}}
\affiliation{Department of Physics, Simon Fraser University, Burnaby, British Columbia, Canada V5A 1S6}

\author{Gong-Bo Zhao}\thanks{E-mail: \url{gbzhao@nao.cas.cn}}
\affiliation{National Astronomical Observatories, Chinese Academy of Sciences, Beijing, 100101, P.R.China}
\affiliation{University of Chinese Academy of Sciences, Beijing, 100049, P.R.China}
\affiliation{Institute for Frontiers in Astronomy and Astrophysics, Beijing Normal University, Beijing, 102206, P.R.China}

\author{Karsten Jedamzik}\thanks{E-mail: \url{karsten.jedamzik@umontpellier.fr}}
\affiliation{Laboratoire de Univers et Particules de Montpellier, UMR5299-CNRS, Universite de Montpellier, 34095 Montpellier, France}

\begin{abstract}
The value of the Hubble constant determined from CMB and BAO measurements is directly dependent on the sound horizon at the photon-baryon decoupling. There has been significant interest in the possibility of new physics at the epoch around recombination that could reduce the sound horizon and increase the inferred value of $H_0$, thus helping to relieve the Hubble tension. One way to determine if new physics is required would be to measure $H_0$ from BAO and CMB without assuming any model for computing the sound horizon. In this study, we use the recently released DESI Year 1 BAO data combined with the CMB acoustic scale and the Planck \lcdm \ prior on $\om$ to determine $H_0$ while treating the sound horizon at baryon decoupling $\rd$ as a free parameter. We find $H_0=69.48 \pm 0.94$ km/s/Mpc, which is $\sim2\sigma$ larger than $H_0 = 67.44 \pm 0.47$ km/s/Mpc in the Planck-best-fit $\Lambda$CDM where $\rd$ is derived using the standard recombination model. For comparison, we perform the same analysis using the pre-DESI BAO data with the CMB acoustic scale and the same prior on $\om$, finding $H_0= 68.05 \pm 0.94$ km/s/Mpc. This difference derives from the notably larger value of the product $\rdh$ measured by DESI. We compare results obtained with and without including the Pantheon Plus sample of uncalibrated supernovae magnitudes in our analysis. Future BAO data from DESI will help determine if the cosmological model at the epoch of recombination model requires a modification.
\end{abstract}


\section{Introduction}

The $\Lambda$ Cold Dark Matter ($\Lambda$CDM) model has been very successful in describing the detailed spectrum of the Cosmic Microwave Background (CMB) anisotropies and measures of cosmic structure formation with just a few cosmological parameters. Notwithstanding this success, the value of the Hubble constant derived from the $\Lambda$CDM best fit to the Planck CMB data, $H_0 = 67.4\pm 0.5$ km/s/Mpc~\citep{Planck:2018vyg}, is lower than that obtained from ``direct'' measurements of the present-day expansion rate, most notably the $H_0 = 73.0 \pm 1.0$ km/s/Mpc value obtained by the Supernovae H0 for the Equation of State (SH0ES) collaboration using Cepheid-calibrated supernovae (SN)~\citep{Riess:2021jrx}. SN-based analyses using alternative calibration methods~\citep{Freedman:2023jcz} also yield higher values of $H_0$, albeit at a lower level of tension with CMB. 

The value of $H_0$ derived from CMB depends on the comoving sound horizon at photon decoupling, $\rstar$, which sets the physical scale for the acoustic features imprinted in the anisotropies. This is closely related to the sound horizon at the baryon decoupling, $\rd$, that sets the characteristic scale of Baryon Acoustic Oscillations (BAO) in the distribution of large scale structure. Both CMB and BAO measure the angular size of the acoustic scale at the respective redshifts, and a smaller $\rd$ would imply a shorter comoving distance to the epoch of decoupling and, thus, a larger $H_0$.

Results from various independent ways of determining $H_0$~\citep{Abdalla:2022yfr} support an interesting trend that measurements which do \emph{not} rely on a model of recombination yield values of $H_0$ in the $69-74$ km/s/Mpc range~\citep{Pesce:2020xfe,Wong:2019kwg,Shajib:2019toy,Harvey:2020lwf,Millon:2019slk}, while estimates that use the standard treatment of recombination give $H_0$ around $67-68$ km/s/Mpc~\citep{Ivanov:2019pdj,Aiola:2020azj,Alam:2020sor}. The trend suggests that there could be a missing ingredient in our current understanding of the universe prior and/or during the epoch of recombination -- an ingredient that could affect the value of the sound horizon at the epoch of photon-baryon decoupling. This motivates identifying combinations of cosmological datasets that can test the validity of the standard model for computing the sound horizon at decoupling.

Assuming the post-recombination expansion history is described by the flat $\Lambda$CDM cosmology, measuring the angular sizes of the acoustic imprints in the distribution of galaxies at different redshifts allows one to determine the product $\rd H_0$ and the current fraction of non-relativistic matter, $\Om$, without making any assumptions about the recombination epoch physics. To break the degeneracy between $\rd$ and $H_0$, further assumptions have to be made. A number of studies ({\it e.g.}~\citealt{Addison:2017fdm}), including the recent analysis of the Dark Energy Spectroscopic Instrument (DESI) Year 1 data~\citep{DESI:2024mwx}, combined the BAO data with the Big Bang Nucleosynthesis (BBN) prior on the baryonic density parameter, $\Omega_bh^2$, which was then used in the standard recombination model to compute $\rd$ and determine $H_0$. Although interesting as a consistency test of some aspects of the $\Lambda$CDM model, this approach does not allow for the possibility of model extensions that predict smaller values of $\rd$. While such extensions may be unable to fully resolve the Hubble tension \citep{Jedamzik:2020zmd}, they can significantly reduce its statistical significance by elevating the Hubble constant derived from the combination of CMB and BAO to $H_0 \sim 70$ km/s/Mpc. It is, therefore, important to find ways to determine $\rd$ and $H_0$ from BAO without assuming a model for computing $\rd$.

In \cite{Pogosian:2020ded}, we proposed treating $\rd$ and $H_0$ as independent parameters and determine their values from BAO combined with a Gaussian prior on the physical matter density parameter $\om$. Such an approach is justified for two reasons: a) in many extensions of the standard model that succeed in reducing the sound horizon, while maintaining a good fit to the CMB spectra, the best-fit value of $\om$ is very close to that of $\Lambda$CDM\footnote{One notable exception is the class of Early Dark Energy (EDE) models~\citep{Poulin:2018cxd}, that require a larger $\om$, which tends to worsen the agreement with the galaxy weak lensing data~\citep{Hill:2020osr,Ivanov:2020ril,DAmico:2020ods,Murgia:2020ryi,Smith:2020rxx,Niedermann:2020qbw}.}; b) one can check if the values of $H_0$ and $\rd$ derived from BAO with the Planck prior on $\om$ agree with the Planck-best-fit \lcdm \ values obtained assuming standard recombination, thus providing a consistency test of the model used to compute $\rd$.

Other determinations of $H_0$ without using a recombination model explored the dependence of the matter power spectrum on the scale set by the horizon at radiation-matter equality. These included combining CMB lensing with uncalibrated supernovae (which provide a measurement of $\Om$)~\citep{Baxter:2020qlr}, combining the matter power spectrum with a prior on $\Om$ from supernovae~\citep{Philcox:2020xbv}, and combining BAO with the CMB and galaxy weak lensing data~\citep{Pogosian:2020ded}. Another possibility is to combine BAO with the cosmic chronometers data~\citep{Pogosian:2020ded,Lin:2021sfs}. The uncertainty in $H_0$ from these measurements is still sufficiently large to make them consistent with both Planck and SH0ES values.

In this paper, we use the DESI Year 1 BAO data~\citep{DESI:2024mwx} along with the Planck measurement of the CMB acoustic scale $\ts$ and the Planck prior on $\om$ to determine the values for $r_d$ and $H_0$. We find $H_0 = 69.48 \pm 0.94$ km/s/Mpc, which is $\sim 2\sigma$ larger than the Planck-best-fit $\Lambda$CDM value of $H_0=67.44 \pm 0.47$ km/s/Mpc. For comparison, we also perform the same analysis using the pre-DESI BAO data, finding $H_0= 68.05 \pm 0.94$ km/s/Mpc. This difference derives from the notably larger value of the product $\rdh$ measured by DESI, $\rdh = 101.89 \pm 1.25$ Mpc, compared to the pre-DESI BAO value of $\rdh=100.53 \pm 1.23$ Mpc. Including the Pantheon Plus sample~\citep{Brout:2022vxf} of uncalibrated supernovae magnitudes, which prefers a larger value of $\Om$, has the expected effect of lowering the BAO-derived values of $H_0$.

In what follows, we describe our methods in Section~\ref{sec:methods} before presenting the results in Section~\ref{sec:results}. We conclude in Section~\ref{sec:summary}.

\section{Methods}
\label{sec:methods}

There are three types of BAO observables~\citep{Eisenstein:2005su}: the angular size of the acoustic feature measured using correlations in the direction perpendicular to the line of sight,
\be
\beta_\perp(z) = D_M(z)/\rd ,
\ee
the same feature measured in the direction parallel to the line of sight, and the angle-averaged or ``isotropic'' measurement. We explain our method using $\beta_\perp$ as our example, but the same arguments also apply to the other two BAO observables. Assuming a flat $\Lambda$CDM cosmology and ignoring the radiation density\footnote{Radiation density is included in our numerical analysis.}, which is negligible at redshifts of interest, $\beta_\perp$ measured at a given redshift $z$ can be written as
\be
\beta_\perp(z) = \int_0^z {2998 \ {\rm Mpc} \ {\rm d}z' \over \rdh \sqrt{\Om(1+z')^3 + 1- \Om}} \ ,
\label{beta_rdh}
\ee
where $h = H_0/(100 {\rm km/s/Mpc})$. From Eq.~(\ref{beta_rdh}) it is clear that having BAO measurements at multiple redshifts allows one to measure two numbers: $\rdh$ and $\Om$. The degeneracy between $\rd$ and $h$ can be broken with a prior on any combination of $\Om$ and $h$, such as a prior on $\om$.

To derive our constraints on $\rd$ and $H_0$, we use {\tt Cobaya}~\citep{Torrado:2020dgo,2019ascl.soft10019T} modified to work with $\rd$ as an independent parameter. The cosmological parameters we vary are $\rd$, $H_0$ and $\om$.

We use the recently published DESI Year 1 (DESI Y1) BAO data~\citep{DESI:2024uvr}, including the BAO distance measurements derived from the DESI Bright Galaxy Sample (BGS, $0.1<z<0.4$) \citep{Hahn:2022dnf}, Luminous Red Galaxy Sample (LRG, $0.4<z<1.1$) \citep{DESI:2022gle}, Emission Line Galaxy Sample (ELG, $1.1<z<1.6$) \citep{Raichoor:2022jab}, Quasar Sample (QSO, $0.8<z<2.1$) \citep{Chaussidon:2022pqg} and the Lyman-$\alpha$ Forest Sample (Ly$\alpha$, $1.77<z<4.16$) \citep{KP6s4-Bault}. Obtaining measurements of different types and over a broad range of redshifts is important, as the intersection of the degenerate bands in the $\Omega_m$-$r_d H_0$ plane corresponding to individual BAO data points yields a precise determination of both quantities. 

The angular size of the CMB acoustic scale,
\be
\ts = \rstar /D_M(z_\star) \ ,
\ee
where $z_\star$ is the redshift of the photon decoupling, can be ``converted'' into a perpendicular BAO observable  $\beta_\star$ as
\be
\bs = {\rstar \over \ts \rd} \ .
\ee
We derive $\bs$ using $100 \ts = 1.04104 \pm 0.00025$, $\rstar = 144.73 \pm 0.23$~Mpc and $\rd = 147.45 \pm 0.23$~Mpc measured by fitting \lcdm \ to the combination of Planck PR4 (NPIPE) HiLLiPoP, LoLLiPoP and CMB lensing likelihoods along with the PR3 Commander low-$\ell$ temperature anisotropy likelihood (in what follows, we will refer to this combination of likelihoods simply as ``Planck''), following the analysis in \citep{Tristram:2023haj}\footnote{The values and the uncertainties of the relevant parameters obtained by fitting \lcdm \ to NPIPE Camspec~\citep{Rosenberg:2022sdy} and PR3 low-$\ell$ EE instead of HiLLiPoP and LoLLiPoP would not make a difference to our analysis as they are in excellent agreement.}. This gives $\rd/\rstar = 1.0188 \pm 0.0024$. As the uncertainty in $\rd/\rstar$ is an order of magnitude larger than that in $\ts$, it dominates the uncertainty in $\bs$, giving 
\be
\bs = 94.286 \pm 0.217 \ ,
\ee
which we use as our ``BAO'' data point at $z_\star=1090$. Note that the comoving distance $D_M(z)$ asymptotes to a constant value at high redshifts and, for all practical purposes, is not sensitive to the precise value of $z_\star$. The ratio $\rd/\rstar$ is also largely the same across different extensions of \lcdm \ model\footnote{The $\rd/\rstar$ ratio is found to be effectively the same across many extensions of \lcdm \ that relieve the Hubble tension by reducing $\rstar$, including additional relativistic degrees of freedom, massive neutrinos, primordial magnetic fields, early dark energy and others. We thank Helena Garc\'\i{}a Escudero for confirming this for the models studied in~\citet{Escudero:2022rbq}.}. Thus, we can use $\bs$ as a ``model-independent'' BAO measurement at $z_\star$. In what follows, we will refer to it as the ``CMB acoustic scale'' measurement.

We supplement our BAO dataset with a Gaussian prior on $\om$ based on the \lcdm \ value measured by Planck, $\om = 0.142 \pm 0.001$. 

Measurements of supernovae Type Ia (SN) luminosities at multiple redshifts, with the intrinsic luminosity unknown, provide an independent measurement of $\Om$, offering an additional consistency test of the \lcdm \  model. Thus, we perform our analysis with and without the Pantheon Plus sample~\citep{Brout:2022vxf} of uncalibrated SN (\sn).

In addition, we also perform our analysis using the BAO data that was available prior to DESI. Our pre-DESI BAO dataset, denoted as \sdss, includes the Date Release (DR) 16 of the extended Baryon Oscillation Spectroscopic Survey (eBOSS) \citep{eBOSS:2020yzd} BAO measurement from LRGs, ELGs, the QSO sample and  the Lyman-$\alpha$ forest, combined with the BOSS DR12 LRG BAO measurements~\citep{BOSS:2016wmc}, low-$z$ BAO measurements by 6dF \citep{Beutler:2011hx} and the SDSS DR7 main Galaxy sample (MGS) \citep{Ross:2014qpa}.

\section{Results}
\label{sec:results}

\begin{table*}[htbp]
\centering
\begin{tabular}{c|c|c|c|c|c}
\hline\hline
        & $\om$ & $r_{\rm d} h$ [Mpc] & $\Om$  & $\rd$ [Mpc] &  $H_0$ [km/s/Mpc] \\
\hline
Planck $\Lambda$CDM & $0.142 \pm 0.001$ & $99.44 \pm 0.82$ & $0.3126 \pm 0.0064$ & $147.44 \pm 0.23$ & $67.44 \pm 0.47$ \\
\sn & - & -  & $0.332 \pm 0.018$ & -   &  -  \\
\sdss BAO & - & $100.53 \pm 1.23$  & $0.298 \pm 0.016$ & -   &  -  \\
\sdss BAO + $\bs$ & - & $100.15 \pm 1.07$  & $0.307 \pm 0.009$ & -   &  -  \\
\sdss BAO + $\om$ & Planck prior &  $100.58 \pm 1.25$  & $0.298 \pm 0.016$ & $145.6 \pm 2.5$   &  $69.08 \pm 1.84$  \\
\sdss BAO + $\bs$ + $\om$ & Planck prior & $100.18 \pm 1.04$  & $0.307 \pm 0.008$ & $147.2 \pm 0.8$   &  $68.05 \pm 0.94$  \\
\sdss BAO + \sn + $\bs$ + $\om$ & Planck prior & $99.59 \pm 0.94$  & $0.312 \pm 0.008$ & $147.5 \pm 0.8$   &  $67.51 \pm 0.86$  \\
DESI Y1 BAO  & - & $101.89 \pm 1.25$  & $0.295 \pm 0.014$ & -   &  -  \\
DESI Y1 BAO + $\bs$ & - & $101.89 \pm 1.01$  & $0.295 \pm 0.008$ & -   &  -  \\
DESI Y1 BAO + $\om$ & Planck prior &  $101.91 \pm 1.25$  & $0.294 \pm 0.014$ & $146.6 \pm 2.1$   &  $69.5 \pm 1.7$  \\
DESI Y1 BAO + $\bs$ + $\om$ & Planck prior & $101.89 \pm 1.02$  & $0.294 \pm 0.008$ & $146.7 \pm 0.8$   &  $69.48 \pm 0.94$  \\
DESI Y1 BAO + \sn + $\bs$ + $\om$ & Planck prior & $101.16 \pm 0.92$  & $0.300 \pm 0.007$ & $147.1 \pm 0.8$   &  $68.78 \pm 0.86$  \\
\hline\hline
\end{tabular}
\caption{\label{tab:params} The mean parameter values and 68\% CL uncertainties derived from the considered combinations of datasets.}
\end{table*}

\begin{figure}[htbp] 
\includegraphics[scale=0.3]{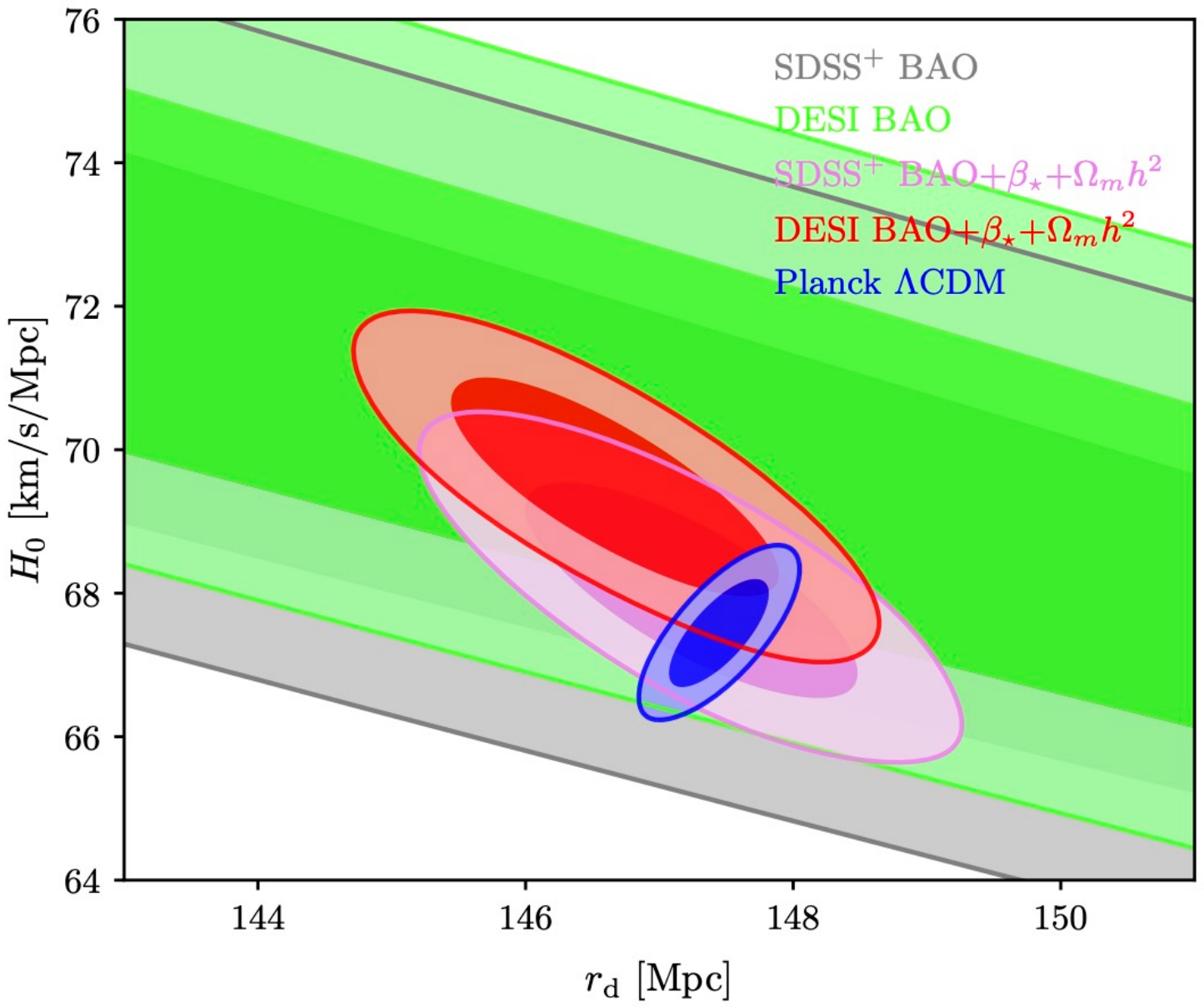}
\caption{Marginalized 68\% and 95\% confidence level CL contours for $H_0$ and $\rd$ from the SDSS$^+$ and DESI Y1 BAO data, and the two BAO datasets combined with the CMB acoustic scale $\bs$ and a prior on $\om$. Shown also are the contours for the best fit Planck \lcdm model.} 
\label{fig1}
\end{figure}

\begin{figure*}[htbp] 
\includegraphics[scale=0.32]{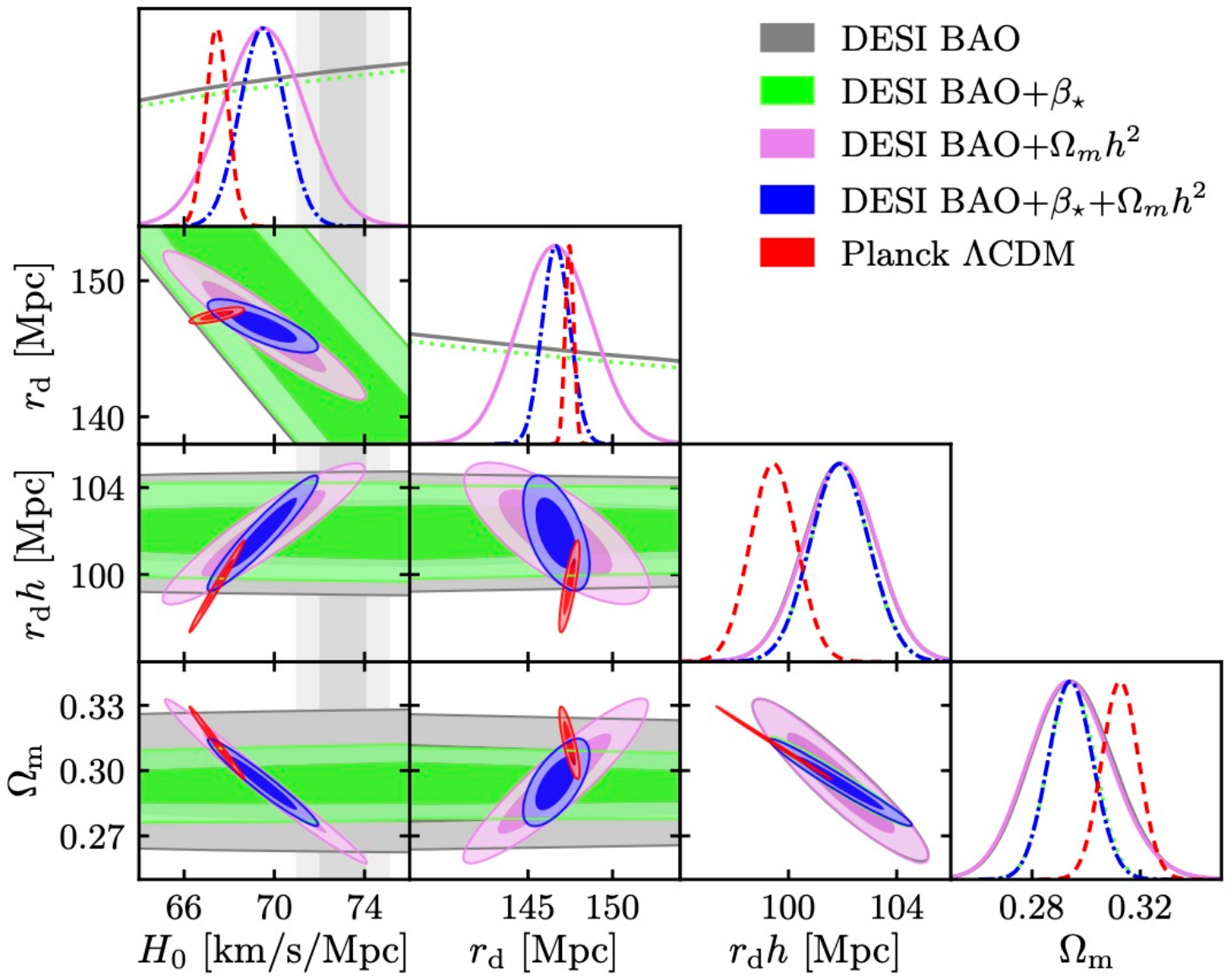}
\includegraphics[scale=0.32]{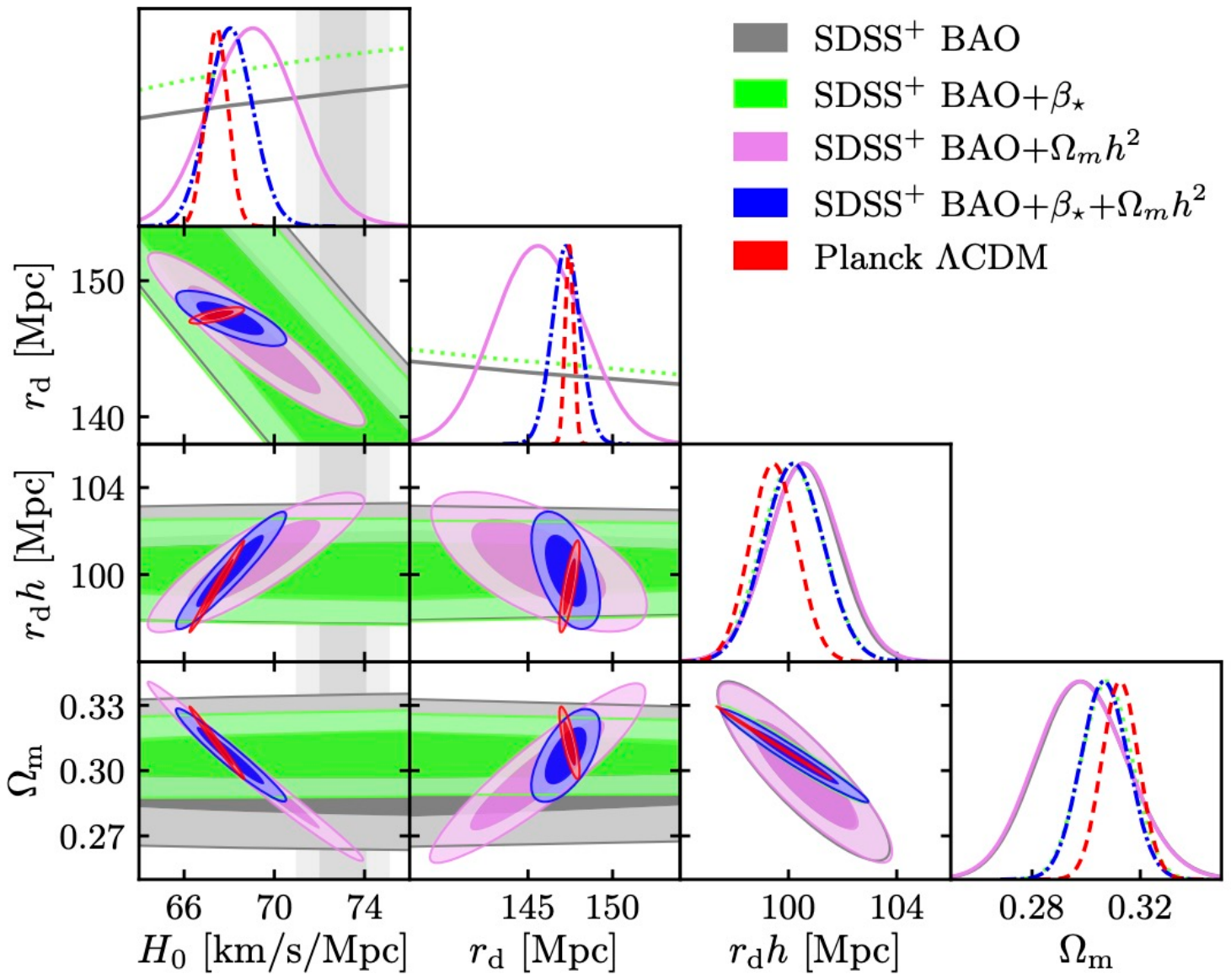}
\caption{{\it Left panel:} constraints on  $H_0$, $\rd$, $\rdh$ and $\Om$ derived from the DESI Y1 BAO data, from DESI Y1 BAO combined with a prior on $\om$, from DESI Y1 BAO combined with the CMB acoustic scale $\bs$ and a prior on $\om$, and the best fit Planck \lcdm. {\it Right panel:} same analysis as in the right panel using the pre-DESI BAO data.} 
\label{fig2}
\end{figure*}

Our results are presented in Table~\ref{tab:params} and Figures~\ref{fig1} and \ref{fig2}. In Fig.~\ref{fig1} we compare the constraints on $H_0$ and $\rd$ from the \sdss \ and DESI Y1 BAO data by themselves and in combination with the CMB acoustic scale $\bs$ and a prior on $\om$. As explained in Sec.~\ref{sec:methods}, BAO data by itself only constrains the product $\rdh$, corresponding a band in the $\rd$-$H_0$ plane. One can see that the DESI Y1 band is shifted upwards relative to the \sdss \ band, corresponding to a larger value of $\rdh$. As shown in Table~\ref{tab:params}, while \sdss \ BAO gives $\rdh = 100.53 \pm 1.23$ Mpc, in excellent agreement with the Planck-best-fit \lcdm \ value of $99.44 \pm 0.82$ Mpc, DESI Y1 yields $\rdh = 101.89 \pm 1.25$ Mpc.

Adding the Gaussian prior on $\om$ and the CMB acoustic scale $\bs$ allows one to constrain both $H_0$ and $\rd$, as also shown in Fig.~\ref{fig1}. For \sdss, this combination yields $H_0=68.05 \pm 0.94$ km/s/Mpc, in good agreement with the Planck-best-fit value of $H_0 = 67.44 \pm 0.47$ km/s/Mpc, while the same combination for DESI Y1 gives a value that is $1.94\sigma$ higher, namely, $H_0=69.48 \pm 0.94$ km/s/Mpc.

Further observations can be made from Table~\ref{tab:params} and Fig.~\ref{fig2}. Without $\bs$, combining the BAO data with the $\om$ prior gives similar results for \sdss \ and DESI Y1, $H_0 = 69.08 \pm 1.84$ and $H_0 = 69.50 \pm 1.70$ km/s/Mpc, respectively. Adding the $\bs$ data point to DESI Y1 halves the uncertainty in $H_0$ while leaving the mean value effectively the same. Doing so for \sdss \ shifts the mean value to the lower end of the $1\sigma$ range. This suggests that the DESI Y1 BAO is in a better consistency with the CMB acoustic scale when one treats $\rd$ as a free parameter, as opposed to computing it using the standard recombination model. The value of $H_0$ inferred in this way, $69.48 \pm 0.94$ km/s/Mpc, is at a $\sim 1.94\sigma$ tension with the Planck-best-fit \lcdm \ value, $67.44 \pm 47$ km/s/Mpc, based on the standard recombination model. On the other hand, \sdss \ is more consistent with the Planck-best-fit \lcdm \ model with standard recombination, but in slight tension with the CMB acoustic peaks if no recombination model assumed.

Both, uncalibrated BAO and uncalibrated SN, provide a measurement of $\Om$. As one can see from Table~\ref{tab:params}, the \sn \ value of $\Om = 0.332 \pm 0.018$ is notably larger than the values obtained from either \sdss or DESI Y1. This difference  indicates a difficulty of fitting the shape of the conformal distance {\it vs} redshift curve obtained from BAO and SN using the same value of $\Om$ parameter. In the context of our analysis, adding the \sn \ data effectively adds a Gaussian prior on $\Om$ which, since $\om$ is constrained, lowers the deduced values of $H_0$. While the parameter values change, the general trends in the differences between \sdss and DESI Y1 remain.

The differences between the \sdss \ and DESI Y1 BAO data and the possible reasons for them have been discussed extensively in~\citep{DESI:2024uvr,DESI:2024mwx} and examining them further is beyond the scope of this paper. There are notable differences in the BAO extracted from the Luminous Red Galaxy samples in the redshift bins centred at $z = 0.5$ and  $z=0.7$. Resolving these differences would likely have to wait until BAO derived from DESI data covering larger volumes is available.

\section{Conclusions}
\label{sec:summary}

The Hubble tension and attempts to resolve it by modifying the physics of (or at) recombination motivate finding ways to determine $H_0$ and the sound horizon at the epoch of baryon decoupling $\rd$ without relying on a recombination model or on the SN calibration. We have analyzed the recent BAO data from DESI Year 1 without using a model to compute $\rd$. We confirmed that DESI Y1 BAO data prefers a larger value of the product $\rdh$ that is in mild tension with the Planck-best-fit $\Lambda$CDM value and the pre-DESI BAO data. Supplementing the BAO data with a prior on $\om$ allows us to measure $H_0$ and $\rd$ while treating the latter as a free parameter, without assuming a recombination model.

Combining DESI Year 1 BAO with the CMB acoustic scale $\bs$ as another "BAO" measurement, and the \lcdm \ Planck prior on $\om$, we find $H_0=69.48 \pm 0.94$ km/s/Mpc. This is $\sim 2\sigma$ away from both the SH0ES and Planck $H_0$ values, while in good agreement with the values of $H_0$ obtained from supernovae calibrated using alternative methods~\citep{Freedman:2023jcz}.  SDSS+ BAO is more consistent with the Planck \lcdm model with standard recombination, but in slight tension with the CMB acoustic scale if no recombination model is assumed. DESI Y1 is less consistent with Planck \lcdm with standard recombination, but in perfect agreement with the CMB acoustic scale if no recombination model is assumed. 

The value of $H_0$ determined from the 5 year DESI BAO data with a prior on $\Omega_mh^2$ and not assuming a recombination model would have an uncertainty that is comparable to the uncertainty in the Planck-best-fit \lcdm~\citep{Pogosian:2020ded}. This will help determine if standard recombination within \lcdm \ is the correct model, or if an amended model with a reduced sound horizon will be required, with the benefit of relieving the Hubble tension.

\acknowledgments

We thank John Peacock for a useful discussion and Helena Garc\'\i{}a Escudero for checking the values of $\rd/\rstar$ in alternative models.
We gratefully acknowledge using {\tt GetDist} \citep{Lewis:2019xzd}. This research was enabled in part by support provided by the BC DRI Group and the Digital Research Alliance of Canada ({\tt alliancecan.ca}). L.P. is supported in part by the National Sciences and Engineering Research Council (NSERC) of Canada. G.B.Z. is supported by the National Key R \& D Program of China (2023YFA1607800, 2023YFA1607803), NSFC grants (No. 11925303 and 11890691), a CAS Project for Young Scientists in Basic Research (No. YSBR-092), a science research grant from the China Manned Space Project with No. CMS-CSST-2021-B01, and by the New Cornerstone Science Foundation through the XPLORER prize.
 

\end{document}